\documentstyle[11pt]{article}

\def\etal{{et al\ }}

\newbox\grsign \setbox\grsign=\hbox{$>$} \newdimen\grdimen \grdimen=\ht\grsign
\newbox\simlessbox \newbox\simgreatbox \newbox\simpropbox
\setbox\simgreatbox=\hbox{\raise.5ex\hbox{$>$}\llap
     {\lower.5ex\hbox{$\sim$}}}\ht1=\grdimen\dp1=0pt
\setbox\simlessbox=\hbox{\raise.5ex\hbox{$<$}\llap
     {\lower.5ex\hbox{$\sim$}}}\ht2=\grdimen\dp2=0pt
\setbox\simpropbox=\hbox{\raise.5ex\hbox{$\propto$}\llap
     {\lower.5ex\hbox{$\sim$}}}\ht2=\grdimen\dp2=0pt

\def\simless{\mathrel{\copy\simlessbox}}

\begin{document}

\begin{center}
\large
{\bf MULTIWAVELENGTH OBSERVATIONS OF \\ 
SHORT TIME-SCALE VARIABILITY IN NGC~4151. \\
III. X-RAY AND GAMMA-RAY OBSERVATIONS}
\end{center}

\normalsize
\vspace{2 cm}


\begin{center}

{\bf R. S. Warwick$^{1}$, D. A. Smith$^{1}$, T .Yaqoob$^{2}$, R.Edelson$^{3}$,
W. N. Johnson$^{4}$,  G. A. Reichert$^{2}$, J. Clavel$^{5}$, 
P. Magdziarz$^{6}$, B.M. Peterson$^{7}$, A. A. Zdziarski$^{8}$} 

\end{center}

\vspace{1 cm}

$^{1}$Department of Physics and Astronomy, University of Leicester,
Leicester, LE1 7RH, UK.\\
$^{2}$NASA Goddard Space Flight Center, Laboratory for High Energy Physics,
Greenbelt, MD 20771, USA.\\
$^{3}$Department of Physics and Astronomy, University of Iowa, Iowa City,
IA 52242, USA. \\
$^{4}$E.O. Hulbert Center for Space Research, Naval Research Laboratory, 
Code 4151, 4555 Overlook SW, Washington, DC 20375-5320, USA. \\
$^{5}$ISO Observatory, ESTEC, Postbus 299, 2200 AG Noordwijk, The 
Netherlands. \\ 
$^{6}$Astronomical Observatory, Jagiellonian University, Orla 171, 
30-244 Cracow, Poland. \\
$^{7}$Department of Astronomy, Ohio State University, Columbus, OH 43210,
USA. \\
$^{8}$N.Copernicus Astronomical Center, Bartycka 18, 00-716 Warsaw, Poland. \\


\vspace{1.0 cm}

\pagebreak

\begin{abstract}

A series of {\sl ROSAT\/}, {\sl ASCA\/} and Compton Gamma-Ray 
Observatory ({\sl CGRO\/}) observations of the Seyfert galaxy NGC 4151
were carried out during the period 1993 November 30 to 
December 13 as part of an intensive campaign to study the
multiwavelength spectral characteristics of its short time-scale
variability. In the softest X-ray bands monitored
by {\sl ROSAT} (0.1--0.4 keV, 0.5--1.0 keV) the source flux remained  
constant throughout the observing period. However, in an 
adjacent band (1.0--2.0 keV) significant variability was evident, the 
most obvious feature being a marked increase (factor 1.45) in the count 
rate over a timescale of $\sim 2$ days commencing roughly 3 days into the 
monitoring period. In contrast, only a low amplitude of variability 
($\simless 10\%$) was measured in the four {\sl ASCA} observations in the 2-10 keV
band (but note that the first {\sl ASCA} observation was performed somewhat 
after the onset of the flux increase seen by {\sl ROSAT}).
The count rates recorded by the Oriented Scintillation
Spectrometer Experiment (OSSE) on  {\sl CGRO\/} are consistent with
$\pm 15\%$ variations in the 50--150 keV gamma-ray band but there is no 
direct correspondence between the gamma-ray and soft X-ray light curves.

The 0.1 to $\sim300$ keV spectrum of NGC 4151 is dominated by a hard
power-law continuum which is cut-off at both high ($\sim90$ keV)
and low ($\sim4$ keV) energy. A high energy cut-off is characteristic of a 
continuum generated by the process of thermal Comptonization whereas that 
at low energy arises from absorption in line-of-sight gas. In NGC 4151
this gas may be partially photoionized by the continuum source but still 
retains significant opacity below 1 keV. The observed soft X-ray 
variability may be the result of changes in the level of the underlying 
soft-hard X-ray continuum or changes in the line-of-sight absorption. 
The data marginally favour the former, in which case the difference 
between the soft X-ray and gamma-ray light curves implies a steepening of the 
continuum as the source brightens, consistent with earlier observations. 
As noted in earlier studies there is a soft excess below 1 keV
which probably arises from more than one scattered and/or thermal
components.

The 1--2 keV soft X-ray and the ultraviolet continuum light curves 
(e.g. near 1440\AA~) show reasonably good correspondence, 
although the relative amplitude of the variations is much higher in the 
X-ray data. The observed ultraviolet to X-ray correlation has
a similar slope to that established in earlier studies, although
a significant residual ultraviolet flux is evident in the recent
observations.  
A possible interpretation is that the X-ray to gamma-ray continuum
is produced in a patchy dissipative corona above the surface of
an accretion disk and that the correlated ultraviolet flux results 
from the reprocessing of part of this continuum by the disk. The 
residual ultraviolet flux may then arise from the reprocessing and/or 
the viscous heating of the disk.

\vspace{1.0cm}

\end{abstract}


\newpage

\section{Introduction}

Over the last twenty years the Seyfert galaxy NGC 4151 has 
been extensively monitored in almost all the accessible wavebands. 
Its high energy spectrum  has been particularly well studied 
for the obvious reason that the active galactic nucleus (AGN)
at the centre of this nearby galaxy is one of the brightest extragalactic
sources in the 2-100 keV band. Thus NGC 4151 has invariably 
been selected as a prime target for each new X-ray and gamma-ray 
mission. All this attention has provided a substantial observational
database but unfortunately many of the properties of this AGN 
remain very unclear. Paradoxically this ``archetypal'' source has rather
anomalous behaviour, for example NGC 4151 has a much harder continuum slope 
than is typical of many Seyfert 1/1.5  galaxies (but see Smith \& Done 1995).

The picture which has emerged is of a relatively low-luminosity AGN
with a highly variable X-ray continuum, $L_X \sim 2-20 \times 10^{42}
\rm~erg~s^{-1}$ (2--10 keV). In the medium energy X-ray band there is
evidence for a steepening of the continuum as the source brightens with
an observed range of the energy spectral index of 0.3--0.7
(Perola et al. 1986; Fiore, Perola \& Romano 1990; Yaqoob \& Warwick 1991;
Yaqoob et al. 1993). The 2--10 keV variability is rather sluggish 
compared to that exhibited by some Seyfert galaxies, with a typical 
flux doubling time-scale of $\sim 0.5-1$ d. In addition there are 
occasional flaring events which occur on time-scales of days to weeks 
(Lawrence 1980; Perola et al. 1986; Yaqoob, Warwick \& Pounds 1989; Yaqoob \& 
Warwick 1991; Yaqoob et al. 1993).
Recently the properties of NGC 4151 in the hard X-ray/soft gamma-ray 
band have been better defined with the availability of qood quality data
from the Oriented Scintillation Spectrometer Experiment (OSSE) 
on the Compton Gamma-Ray Observatory ({\sl CGRO\/}). Zdziarski,
Johnson \& Magdziarz (1995) find the OSSE spectral data can be well described
in terms of thermal Comptonization in a source having an intrinsic energy
spectral index of 0.4--0.7 and a plasma temperature of $\sim 40-50$ keV.
Interestingly the 1991-1993 OSSE observations show the gamma-ray spectrum 
to have been almost constant in terms of the total flux and shape 
(Zdziarski, Johnson \& Magdziarz 1995); the true nature of the wide-band 
spectral variability at present remains unclear.
 
The hard X-ray continuum of NGC 4151 is cut-off at low energy 
(below $\sim4$ keV) due to photo-electric absorption in  
line-of-sight gas with  hydrogen
column density varying between $\sim 3$--$15 \times 10^{22}$ cm$^{-2}$
(Yaqoob \& Warwick 1991; Yaqoob et al. 1993).  An observation by the
Solid-State Spectrometer abroad the {\sl Einstein\/} observatory
first showed the soft X-ray attenuation in the NGC 4151 spectrum to be
incompatible with a uniform screen of cold solar abundance gas
(Holt et al. 1980). The observed ``soft excess'' has subsequently been 
described in terms of both
partial covering, a warm absorber and hidden spectral 
components  (Holt et al. 1980; Yaqoob \& Warwick 1991; 
Weaver et al. 1994a,b; Warwick, Done \& Smith 1995).  
There is also a substantial flux observed below 1 keV which may arise,
at least in part, from the spatial resolved soft X-ray source which 
coincides with the Extended Narrow-Line Region in NGC 4151
(Elvis, Briel \& Henry 1983; Elvis et al. 1990; Morse et al. 1995).
A final feature of the X-ray spectrum of NGC 4151 is
an intense iron K$_{\alpha}$ emission line produced by the fluorescence of 
gas illuminated by the continuum X-rays. Recent
{\sl ASCA} spectra of NGC 4151 reveal the presence of both a narrow 
component and a redshifted broad component of this line with
the latter possibly arising close to
the central black hole in a putative accretion disk structure
(see Yaqoob et al. 1995 for further details). 

Here we report the results of a series of observations carried out by
{\sl ROSAT\/}, {\sl ASCA\/} and {\sl CGRO} OSSE, which form part of
an intensive campaign to study the short-timescale 
variability. Paper I in this series (Crenshaw et al. 1995)
reports the ultraviolet observations performed by the {\sl International
Ultraviolet Explorer} ({\sl IUE}), which serve as the reference
measurements for the whole programme. Paper II (Kaspi et al. 1995)
similarly presents the results of the optical monitoring campaign.
The current work (Paper III) concentrates on the high energy observations. 
Finally, Paper IV (Edelson et al. 1995) presents a detailed study of the 
short time-scale variability exhibited by NGC 4151 with emphasis on 
the multiwavelength characteristics of the phenomenon.

\section{The Observations}

A series of {\sl ROSAT\/}, {\sl ASCA\/} and {\sl CGRO\/} OSSE
observations of NGC 4151 were carried out during the period 1993 November
30 to December 13, as part of an intensive multiwavelength
monitoring campaign.  Details of the observing programs of the three
satellites are summarised in Tables 1--3.

\subsection{The {\sl ROSAT\/} Observations}

A total of 13 {\sl ROSAT\/} observations were performed during the
monitoring period, all with the Position Sensitive Proportional
Counter (PSPC) in the focal plane of the X-ray telescope. A standard
operational mode was employed in which the telescope's pointing
direction is ``wobbled'' during the observation (by $\pm 3$ arcmin on
a time-scale of $\sim 400$ seconds) in order to avoid the possibility
of severe shadowing of an X-ray source by the coarse wire mesh which
supports the PSPC window. The background count rate in the PSPC is
relatively low and arises primarily due to cosmic-ray particles,
scattered solar X-rays and the cosmic diffuse X-ray background
(Snowden et al. 1992). For the present work we have applied a data
rejection threshold of 170 count s$^{-1}$ on the high energy particle
detection rate monitored by the PSPC (the so-called Master Veto Rate)
as a simple method of excluding time intervals when the particle
background was unduly high.  The resulting exposure times for the {\sl
ROSAT\/} observations are listed in Table 1. PSPC light curves and
spectra for NGC 4151 and the BL Lac object 1E1207.9+3945 were
extracted from the data using a source cell of radius 2.5 arcmin (the
BL Lac object is located 5 arcmin to the north-west of NGC 4151 and
serves as useful comparison source in the {\sl ROSAT\/} observations).
A corresponding background spectrum was taken from an annular region
of radius 9--15 arcmin centred on NGC 4151 but from which discrete
sources had been excised.  After applying corrections for vignetting
effects, the relative size of the background cell and the effective
exposure, a background-subtracted PSPC spectrum was determined for
each individual observation. Light curves were
derived from the count rates in three spectral bands (corresponding to
the energy ranges 0.1--0.4 keV, 0.5--1.0 keV and 1.0--2.0 keV),
whereas for the spectral analysis the PSPC data were binned into 27
spectral channels covering the full 0.1--2.0 keV energy range (this
represents a significant over-sampling of the spectra in comparison to
the energy resolution of the PSPC).
The {\sl ROSAT\/} PSPC spectra for NGC 4151 were analysed using a
standard model fitting procedure ({\sc xspec\/} version 8.50)
and the DRM31 instrument response matrix (which is
currently recommended for observations performed in the {\sl ROSAT\/}
AO-3 period). A 1\% systematic error was added to the data to take
account of the uncertainties in the response matrix.

\subsection{The {\sl ASCA\/} Observations}

Four observations of NGC 4151 were carried out with the two
Solid-State Imaging Spectrometers (SIS-0 and SIS-1)
and two Gas Imaging Spectrometers (GIS-2 and GIS-3) on 
{\sl ASCA\/} (Tanaka, Inoue \& Holt 1994) as part of the
monitoring campaign (see Table 2). The SIS data, which were taken in 
2-CCD FAINT and BRIGHT modes, were first all converted to BRIGHT mode
and then events of grade 0, 2, 3 and 4 were used in the subsequent analysis. 
Events were accumulated from
circular regions centred on NGC 4151 with extraction radii of 
$3.9^{\prime}$ and
$3.4^{\prime}$ for SIS-0 and SIS-1 respectively, and $6.0^{\prime}$
for GIS-2 and GIS-3. Event lists were also obtained for the off-source
regions. For the SIS, this included all events from the source chip
farther than $6.0^{\prime}$ from NGC 4151 and, for the GIS, from an
annulus of radius $9^{\prime}$--$13^{\prime}$ centred on NGC 4151
(events from the BL Lac were excluded). Both the source and background
lists were examined for anomalous events which were then
rejected; this is a very effective way of removing large numbers of
``flickering'' pixels and light-leakage events. Events recorded during 
spacecraft passages through the South Atlantic Anomaly, during periods
when the magnetic cut-off rigidity was less than $7.5$ GeV/$\rm c^{2}$ 
and at times when the Earth elevation angle was less than $5^{\circ}$ were
also rejected. Finally, any remaining ``hot'' pixels were removed by
examining the count/pixel distributions in a large number of annular
rings.  X-ray spectra were made from the resulting cleaned source and 
background event lists for each detector corresponding to the four 
observing periods. Summed spectra
corresponding to the entire observing period were also
produced. As a check the analysis was repeated with more severe selection 
criteria but this resulted in no significant
improvement in the quality of spectra obtained. For the
purpose of the present paper, we have analysed only the SIS-0 data. These
are the ``best'' data in the sense that SIS-0 is better calibrated and
the source appears closer to the XRT optical axis, giving higher
count rates. There is, in fact, very good agreement amongst the 
four {\sl ASCA\/} instruments for this set of observations. 
Yaqoob et al. (1995) also show that the SIS data require no correction to the
energy scale. The present observations were made shortly after the end of 
the Performance Verification Phase at which point the degradation of the 
CCD charge transfer efficiency had no measurable impact on the data quality. 
Fluctuations in the ``zero'' level of the energy scale by 10--30 eV 
(Dark Frame Error) cause a degradation of the
energy resolution rather than a measurable non-zero offset.

\subsection{The {\sl CGRO\/} OSSE Observations}

{\sl CGRO\/} OSSE  monitored NGC 4151 
continuously over a period of 13 days during the campaign (see Table 3).
OSSE observations consist of a sequence of two-minute integrations on the
source field alternated with similar offset-pointed background measurements.  
Background estimation is achieved by performing a quadratic fit to 
three or four contiguous two-minute background accumulations of 
satisfactory data quality taken from the immediate temporal vicinity 
of each two-minute on-source accumulation. The resulting two-minute 
background-subtracted spectra are then summed over the period of a day.
In the present analysis the data processing and screening produced a total 
on-source livetime of $4.9 \times 10^5$~detector-seconds.  A similar duration
of offset-pointed background observations were used in creating the
background-subtracted spectra. (See Johnson et al. 1993 for a more detailed
description of OSSE performance and analysis.)
The source was detected at a significance level of greater than 
12$\sigma$ per day in the energy
range $50 - 150$~keV.  The daily average flux in this band is seen to vary by
$\pm 15$\% during the observation.  
The average spectrum of NGC 4151 for the whole observing period was created 
by summing the daily spectra for all four detectors.  
For analysis by {\sc xspec\/} the OSSE data were binned into 14 spectral channels 
covering the energy range 50 keV to 1 MeV. Systematic errors 
were added to the statistical errors before the spectral fitting.
These systematics were estimated from uncertainties in the low energy 
calibration and in the detector response and are significant below 100 keV 
with a peak of $\sim 10\%$ of the statistical error at 50 keV.

\section{The {\sl ROSAT} results}

Figure 1 shows the PSPC count rates measured in 3 spectral bands for
both NGC 4151 and the nearby BL Lac object, 1E1207.9+3945, in each of
the 13 {\sl ROSAT\/} observations. In the case of 1E1207.9+3945 there
is no clear evidence for variability over the 10-day monitoring
period. Thus, a $\chi^{2}$ test based on the hypothesis that the BL
Lac source flux is constant gave $\chi^{2}$ values of 7.0, 12.3 and
18.1 for 12 degrees-of-freedom (d-o-f) for the soft, medium and hard
{\sl ROSAT\/} bands respectively.  A similar situation pertains for
the soft and medium band light curves of NGC 4151 (the constant source
hypothesis gave $\chi^{2} = 6.8$ and 6.6 with 12 d-o-f for the two
bands respectively).  However, in the 1.0--2.0 keV {\sl ROSAT} band variability
is evident ($\chi^{2} = 215$ for 12 d-o-f) in the form of a marked
change in the count rate commencing roughly 3 days into the monitoring
campaign at which point the signal increases by about a factor 2
(minimum to maximum) over a period of two days. Although, not strictly
consistent with a constant count rate, the first six {\sl ROSAT\/}
observations (hereafter group A) have a mean count rate in the hard
band of 0.22 count s$^{-1}$, whereas the last five observations (group
B) have a mean count rate of 0.32 count s$^{-1}$; this corresponds to
a 45\% stepwise flux increase.


Our preliminary analysis of the spectral variability exhibited by NGC
4151 is based on the combined datasets, defined as group A and B
above. The disparate light curves of NGC 4151 in the soft/medium and
hard {\sl ROSAT\/} bands (Figure 1) provide clear evidence for a
spectral change. This point is further illustrated in Figure 2, which
shows the ratio of pulse height spectra for the group A and B
datasets; the variability is clearly restricted to spectral channels
above $\sim 1 $ keV. In order to investigate the nature of this soft
X-ray variability we have employed a simple phenomenological model
with two emission components, namely a thermal bremsstrahlung
continuum representative of the bulk of the soft X-ray flux and a hard
power-law continuum which dominates above $\sim 1$ keV. The former
component is presumed to suffer only light absorption due to a modest
line-of-sight column density, $N_{Hgal}$, which arises in our own
galaxy, whereas the latter is relatively heavily absorbed by a gas
column, $N_{H}$ \footnote{Throughout this paper we have used the element 
abundances and absorption cross-sections adopted by Morrison \& McCammon 
(1983)}, intrinsic to NGC 4151 (and presumably located 
in its active nucleus). 

In NGC 4151 there is evidence for a spectral index--flux correlation
in the sense that the X-ray spectrum softens as the source brightens. 
The energy index, $\alpha$, varies between 0.3 and 0.7 at which point 
the correlation appears to ``saturate'' 
(Yaqoob \& Warwick 1991; Yaqoob et al. 1993).  As we
do not know the energy index {\it a priori\/} (the {\sl ROSAT\/} PSPC
has insufficient spectral resolution and bandwidth to
constrain this parameter usefully), in the present analysis we 
have fixed the value of $\alpha$ at 0.5. The
spectral modelling procedure we have adopted is to fit simultaneously
the PSPC spectra from the group A and B datasets with the temperature,
$kT_{brem}$, normalisation, $A_{brem}$, and absorption, $N_{Hgal}$, of
the thermal bremsstrahlung component constrained to the same values for the
two datasets. In contrast, the two free parameters defining the
power-law continuum (normalisation, $A$, and absorption, $N_{H}$) 
were not tied across the two
datasets, thus catering for the variability evident above 1 keV. The
results from the simultaneous fit to the group A and B datasets
(involving a total of 7 free parameters) are given in Table 4. 
(The errors quoted in Table 4 and elsewhere in this paper represent 90 percent 
confidence limits for one interesting parameter, see Lampton, Margon 
\& Bowyer, 1976). From
these results it may be concluded that the gross variability in the
{\sl ROSAT\/} band is attributable mainly to a change in the
normalisation of the hard power-law continuum (by a factor of $\sim
2.4 $) rather than to a variation in the absorption parameter,
$N_{H}$. When the value of $N_{H}$ was tied across the two datasets,
the minimum $\chi^{2}$ for the joint fits showed little variation
compared to that quoted in Table 4.
To further illustrate the influence of variations in the level of the
hard continuum and/or column density changes, we have determined the
variation in the minimum $\chi^{2}$ for models in which the values of
A and $N_{H}$ were constrained to fixed ratios across the group A and
group B datasets.  The result, shown in Figure 3, emphasises the point
that although column density variations cannot be excluded (of magnitude
$ \simless 20\%$) a significant change in the normalisation of the
hard power-law is definitely required by the data. Thus, here
we take the view that the variability revealed by the
{\sl ROSAT\/} observations is largely attributable to variations in
{\it the level of the hard X-ray continuum} in NGC 4151.

The two component spectral model discussed above was also applied to
the spectra from each of the individual {\sl ROSAT\/} observations but
with only the normalisation of the hard power-law considered as a free
parameter (the other parameters of the model being fixed at their
global best-fit values). These spectral fits were then used to derive
the corresponding X-ray fluxes in the 1--2 keV band. The results are
tabulated in Table 1 and also plotted as a light curve in Figure 4. Note
that in the model discussed above the contribution to the flux in the 
1--2 keV band from the (non-variable) thermal bremsstrahlung component 
is $9.2 \times 10^{-13}~\rm~ erg~cm^{-2}~s^{-1}$.

A further point concerns the best fitting parameters for the thermal
bremsstrahlung component. The values quoted in Table 4 are in fair
agreement (within the errors) with those obtained in June 1991 in an
earlier {\sl ROSAT\/} PSPC observation (Warwick et al. 1995); 
however, the observed 
0.2--1.0 keV flux of $\sim 4.9 \times 10^{-12}$ erg cm$^{-2}$ s$^{-1}$
is approximately 30\% higher than that measured
previously. We note that recent {\sl ASCA\/} observations (Weaver et al.
1994b) have provided
evidence for variability (over a time-scale of $\sim 6$ months) in the
the soft X-ray spectrum of NGC 4151 below 1 keV. Clearly such variability
gives an important clue as to the origin of the soft X-ray flux in
NGC 4151 and can be used to set limits on what fraction of the emission 
originates in extended thermal or electron scattered
components. However, at present we are cautious in ascribing the
apparent variations in the sub-1 keV PSPC flux (over a time-scale of
2.5 years) to intrinsic source variability in NGC 4151 since PSPC
calibration uncertainties (particularly in the 
0.1--0.4 keV band) may account, at least in part, for the differences in the
flux measurements.


Finally we note that in the above analysis the selection of
a different energy index other than $\alpha = 0.5 $ but in the range 
0.3--0.7 has only a marginal effect on the spectral fitting results.
The conclusions are essentially unaltered for any value of $\alpha$
within this specified range.

\section{The {\sl ASCA\/} Results}

Comparison of the count rates measured on NGC 4151 from the SIS-0 
telescope  during the four {\sl ASCA\/} observations 
provides evidence for low amplitude ($\simless 10\%$) but 
significant variability above 1 keV. Unfortunately the first {\sl
ASCA\/} observation was performed on 1993 December 4, somewhat after
the onset of the largest amplitude change seen by {\sl ROSAT\/}.

Our preliminary analysis of the {\sl ASCA\/} SIS-0 spectra for the
energy band 0.4--10 keV is based on a partial covering description of 
the complex absorption feature which dominates the 1--5 keV spectrum of 
NGC 4151. Specifically we use the ``dual-absorber'' discussed by Weaver 
et al. (1994b) in which a fraction $C_{F}$ of
the hard X-ray flux suffers absorption in a column density, $N_{H1}$,
with the remaining $1-C_{F}$ fraction subject to a somewhat lower (but
still substantial) column density, $N_{H2}$. Further 
parameters are then required to represent the iron-K line and 
any additional iron edge features\footnote 
{Iron-K edge features arising from a solar abundance concentration 
of iron atoms in the gas columns $N_{H1}$ and $N_{H2}$ are automatically
included in the spectral fitting.} 
apparent in the energy range 
5--8 keV. For simplicity the present analysis only includes 
a narrow iron $\rm K_{\alpha}$ line with an intrinsic width
$\sigma_{\rm K\alpha} = 0.07$ keV (see Yaqoob \etal 1995 for a more detailed 
discussion of the iron features in {\sl ASCA\/} spectra of NGC 4151). 
In addition the bulk of the soft X-ray flux below 1 keV was
initially modelled as a thermal bremsstrahlung
continuum absorbed by a column density $N_{Hgal}$ (c.f. section 3).
At this point $N_{Hgal}$ was fixed at $2.1 \times 10^{20}$
cm$^{-2}$ which is the Galactic HI column density 
measured in the direction of NGC 4151 (and is also reasonably consistent
with the value for $N_{Hgal}$ obtained in the earlier 
{\sl ROSAT\/} analysis).  
As in the previous section we adopt the value $\alpha = 0.5$ for the energy 
index of the hard continuum.

Initially we considered a simultaneous fit to the four {\sl ASCA\/} spectra 
with only the normalisation of the hard power-law continuum, A, free to 
vary across the four datasets; details of the best-fitting model are given 
in Table 5.  This analysis shows that the level of variability in the
underlying X-ray continuum, as measured by its normalisation,
was $\simless 10\%$. 
The minimum $\chi^{2}$ of 1371 (1052 d-o-f) 
demonstrates that this simplified partial covering model
does not provide a full description of the {\sl ASCA\/} spectra. 
There is evidence for additional spectral features and subtle
spectral variability
across the four {\sl ASCA} datasets. This is illustrated in Figure 5 
which shows the ratio of the four {\sl ASCA\/} spectra to the 
best-fitting spectral form ({\it i.e.} the model defined in Table 5). 
Residual features are apparent in all the spectra; for example, 
three out of the four {\sl ASCA\/} spectra show evidence for a narrow line
feature at $\sim 0.89$ keV with an average flux of $0.8 \times 10^{-4}
\rm~photon~s^{-1}~cm^{-2}$ and an equivalent width of $\sim 30$ eV. 
(A similar line feature was noted in previous {\sl ASCA\/} observations by 
Weaver et al. 1994b).  However, across the full 0.5--10 keV band 
there is no clear evidence for systematic spectral variations. 
It is possible that some spectral variability results from changes 
in the complex absorber. To investigate this
we have carried out a spectral fitting analysis with the parameters 
$N_{H1}$, $C_{F}$ and $N_{H2}$, as well as the continuum normalisation,
free to vary across the individual observations. The result was a significant
reduction in the minimum $\chi^{2}$ by 69 to 1302, with 1043 d-o-f;
however, the derived error ranges on the individual parameters were 
large and in all cases consistent with the average parameter values  
obtained earlier (Table 5).  We note that subtle changes in the form 
of the absorption on a timescale of days have been reported 
in previous studies of NGC 4151 (e.g. Weaver et al. 1994a; Warwick et al. 
1995).
 
The derived fluxes from the {\sl ASCA\/} observations in the 1--2 and
2--10 keV energy bands are listed in Table 2 and also plotted in
Figure 4. An important point to note is that there is rather good
agreement between the {\sl ROSAT\/} and {\sl ASCA\/} measurements of
the 1--2 keV flux for those observations which overlap in
time. Significant variability is apparent in the combined {\sl
ROSAT\/}--{\sl ASCA\/} light curve. Unfortunately the coverage of the
2--10 keV band in the {\sl ASCA\/} (only) measurements is too sparse
to confirm the large amplitude variability as a broad-band effect.

\section{The {\sl CGRO\/} OSSE results}

The light curve in the 50--150 keV band derived from the {\sl CGRO\/} OSSE 
observations is shown in Figure 6. These data are {\it
inconsistent\/} with a constant count rate ($\chi^{2} = 28.4$ for 11
d-o-f.) with variations from the mean of up to $\pm 15\%$ peak-to-peak
($\sim 10\%$ rms). Similar
variations have been reported by Maisack et al. (1993) from a previous
OSSE observation of this source. The present gamma-ray
measurements fail to show any significant variation corresponding to the flux
increase apparent in the 1--2 keV {\sl ROSAT\/} band. Interpreting
the latter as a factor $\sim 2.4$ increase in the power-law continuum at
$\sim 1$ keV (section 3) and the former as a continuum increase
at $\sim 100$ keV of no more than $15\%$ (estimated for the relevant
periods from Figure 6) implies a change in the spectral index of
the underlying hard power-law continuum $\Delta \alpha \approx 0.15$. 
We note that this is compatible with the spectral softening as the 2--10 keV 
flux increases which has previously been reported for NGC 4151 
(Perola et al. 1986; Fiore, Perola \& Romano 1990; 
Yaqoob \& Warwick 1991; Yaqoob et al. 1993).

There is no evidence for spectral variability within the set of OSSE 
observations, thus we have considered only the mean spectrum.
This is not well described by a single power-law
continuum (spectral fitting of a two parameter model gave a minimum
$\chi^{2} = 42.6$ for 12 d-o-f.). With this model the spectral index,
$\alpha = 1.61 \pm 0.08$, is much softer than that observed at lower
energies, indicating that the spectrum must break at an intermediate
energy. Thus, we have adopted a spectral model in which a hard power-law 
continuum is modified by an exponential cut-off of the form 
$\exp(-E/E_{\rm c})$, where $E_{\rm c}$ is the characteristic energy. 
If we fix the hard power-law slope 
at $\alpha = 0.5$ (see Section 3 and 4) then $E_{\rm c} = 87^{+8}_{-7}$ keV 
and $\chi^{2} = 7.9$ (12 d-o-f.). In this case the
best-fit normalisation (at 1 keV) is $7.1 \pm 0.7 \times 10^{-2}$ 
photon cm$^{-2}$ s$^{-1}$ keV$^{-1}$, in fairly close agreement with 
the values derived from the fits to the {\sl ASCA\/} data (see Table 5). 
Also, the
observed 50--150 keV flux values for each of the individual OSSE
observations were calculated on the basis of this spectral model; these
are listed in Table 3. 

Alternatively the OSSE data can be fitted with the thermal Comptonization
model of Titarchuk \& Mastichiadis (1994). We obtained best fit parameter
values of $kT = 50^{+25}_{-12}$ keV and $\alpha = 0.65^{+0.21}_{-0.18}$, 
where $\alpha$ is the asymptotic low-energy power-law index (we use 
this parameter rather than the geometry-dependent Thomson optical depth). 
In this case the $\chi^{2}$ was 7.0 for 11 d-o-f. Very similar
results have been obtained by Zdziarski, Johnson \& Magdziarz (1995)
who provide a detailed critique of the Titarchuk \& Mastichiadis model
and also survey the available set of OSSE and other hard X-ray/
gamma-ray observations of NGC 4151.

\section{The wide band X-ray to gamma-ray spectrum of NGC 4151}

Spectral coverage from 0.1 keV to $\sim300$ keV is obtained from the
combined {\sl ROSAT\/}, {\sl ASCA\/} and {\sl CGRO\/} OSSE data sets. 
Here we consider only the first {\sl ASCA\/} observation together with 
the mean spectrum from the near simultaneous {\sl ROSAT\/} observations 
(observations 8 and 9). However, we use the full set of 
{\sl CGRO\/} OSSE 
observations since there is no convincing evidence for spectral variability 
in these data.

The quasi-simultaneous data were initially fitted with the partial
covering model described in Section 4 with the addition of
the exponential cut-off to the hard power-law continuum (Section 5). 
The spectral index was again fixed at $\alpha = 0.5 $.
Table 6 gives the details of the best-fitting parameters
(the iron $\rm K_{\alpha}$ line parameters are not quoted since the values are 
little changed from Table 5). This fit to the combined data demonstrates
that there is reasonable agreement between the three instruments 
as illustrated in Figure 7a, which shows the measured count rate spectra 
compared with the best-fit model.  The excellent correspondence
between the {\sl ROSAT\/} PSPC and {\sl ASCA\/} SIS-0 data in the 
1--2 keV range suggests that the relative cross-calibration of the two
instruments, albeit in this limited energy range, is good to better than 10\% 
(thus validating the combined {\sl ROSAT\/}/{\sl ASCA\/}
1--2 keV light curve in Figure 4). However, below 1 keV, there is
evidence from the spectral fitting residuals for systematic differences 
between the {\sl ROSAT\/} and {\sl ASCA\/} datasets, which presumably 
relate to calibration uncertainties (see Figure 7b). 

The derived input spectrum corresponding to the above partial covering model 
is shown in Figure 7c. This figure illustrates the main characteristic
of the soft X-ray to gamma-ray spectrum of NGC 4151, namely its domination
by a hard power-law continuum which is cut-off at both high 
($\sim 90$ keV) and low ($\sim4$ keV) energy. The high energy cut-off 
is characteristic 
of a continuum generated by the process of thermal Comptonization (see 
section 5; Titarchuk \& Mastichiadis 1994; Zdziarski, Johnson \& 
Magdziarz 1995) whereas that at low energy arises from absorption 
in a substantial
column density ($\sim 10^{23}\rm~cm^{-2}$) of relatively cool gas in
the line-of-sight to the continuum X-ray source. As noted in many earlier 
studies (and in previous sections) there is also a soft excess flux
below 1 keV which is probably due to the superposition of a number
of seperate spectral components (although so far we have used only a
single temperature thermal bremsstrahlung representation). In the remainder
of this section we comment on various aspects of the high energy spectrum 
of NGC 4151.

\subsection{The soft excess flux below 1 keV}  

The thermal bremsstrahlung
description  of the soft X-ray spectrum of NGC 4151 is not satisfactory 
since line emission dominates the cooling of thermal plasmas in the 
temperature range under consideration. In addition the steep slope of the 
0.1--1.0 keV spectrum of NGC 4151 rules out an origin for this flux 
solely as electron scattering of the hard X-ray continuum. To illustrate
this last point we have repeated the above spectral fitting with
the thermal bremsstrahlung component replaced by a soft power-law 
component (with energy index, $\alpha_{soft}$ and normalisation $A_{soft}$).
The results detailed in Table 6 show that
$\alpha_{soft} \approx 1.3$, which is a much steeper slope 
than is measured for the underlying X-ray continuum 
(albeit at energies well above 1 keV). A plausible physical description 
of the 0.1--1.0 keV spectrum of NGC 4151 requires at least two spectral 
components. Comparison of Tables 4 and 5 shows that the 
{\sl ROSAT\/} data alone require a lower temperature for the thermal 
bremsstrahlung component than when fitting the
{\sl ASCA\/} data alone ({\it i.e.} $\sim 0.4$ keV as opposed to $\sim 1.3$
keV). This suggests that an additional ``ultra-soft'' component is
required by the data as previously argued by Weaver et al. (1994b) and 
Warwick et al. (1995).

The {\sl ASCA\/} data are not compatible with a sizeable fraction of
the 0.5--1.0 keV soft X-ray flux originating as thermal emission in a
solar abundance gas (this is based on the lack of prominent O VII, O
VIII and Fe-L emission lines). However, it is plausible that the 
bulk of the flux observed in this narrow spectral range is predominantly the
extrapolated hard X-ray continuum seen as a result of pure electron
scattering in a medium extending along the symmetry axis of the source
(by analogy with the situation pertaining in NGC 1068, see Marshall et al.
1993 and references therein). 
Thus, a possible two component model involves a power-law continuum 
(with spectral index similar to that of the hard power-law continuum) 
plus an ultra-soft component. The latter may be modelled as either 
a $\sim 10^{6}$ K thermal plasma (perhaps
associated with the extended source close to the nucleus of NGC 4151;
Elvis et al. 1983; Morse et al. 1995) or a scattered black-body
component (possibly originating in the hot inner regions of a putative
accretion disc). For a more detailed discussion see Weaver et al. (1994a,b) 
and Warwick et al. (1995).

\subsection{Complex absorption in the 1--5 keV band} 

Although partial covering 
provides a viable description of the complex absorption observed in 
the 1--5 keV spectrum of NGC 4151, an alternative and perhaps more 
physically realistic approach is to utilise the reduced soft X-ray opacity 
exhibited by a partially ionized medium. This 
``warm absorber'' model has previously been considered in the NGC 4151 
context by a number of authors (Yaqoob \& Warwick 1991; Netzer 1993; 
Weaver et al. 1994a,b; Warwick et al. 1995).
In modelling the composite {\sl ROSAT\/}--{\sl ASCA\/}--OSSE spectrum
with a warm absorber we replace the three partial covering
parameters with just two, namely the column density of the medium
$N_{H}$ and the ionization parameter $\xi$ (where $\xi =
L_{ion}/nr^{2}$ and the ionizing luminosity, $L_{ion}$, is obtained by
integrating the extrapolated continuum between 5 eV and 300 keV; 
see Warwick et al. 1995 and Zdziarski et al. 1995). 

We initially revert to the case where the only additional soft component is
the scattered power-law mentioned above (with the spectral index set 
equal to that of the hard X-ray continuum). With this model
a reasonable fit was obtained  ($\chi^{2} = 412$/ 302 d-o-f)
with $N_{H} = 9.6 \times 10^{22}~\rm~cm^{-2}$ and $\xi = 28$; 
the corresponding input model spectrum is shown in Figure 8a.
However, there is an immediate difficulty with this particular description.
>From the {\sl ROSAT\/} observations it is clear that
there was no variability below $\sim 1$ keV at the time the hard X-ray
continuum varied substantially. This essentially rules out the situation
(as in Figure 8a) in which a significant fraction of the 0.1--1.0 keV
flux arises due to the leakage of the underlying X-ray continuum through the
warm absorbing medium (in which there is reduced soft X-ray opacity 
as a result of photoionization, e.g. Krolik \& Kallman 1987), 
since in such a model the ultra-soft flux must track the continuum variation. 
This problem can be circumvented by invoking a somewhat lower value for the 
ionization parameter or, alternatively, including an additional (cold) 
column density to suppress the transmitted soft X-ray flux. Either way, at
least one additional soft spectral component is required to account
for the ultra-soft excess flux. By way of illustration, Figure 8b shows
the input spectrum for a revised warm absorber model (which gave
$\chi^{2} = 396$/300 d-o-f) where $N_{H} = 8.2 \times 10^{22}~\rm~cm^{-2}$ 
and $\xi = 14$ and which includes an additional 
$\sim 10^{6}$ K Raymond Smith thermal component (with solar element
abundances assumed). The decrease in 
the ionization parameter by about a factor of two, greatly reduces the 
leakage of the hard continuum below 1 keV with the deficit being made up,
of course, by the thermal emission.  Thus in this setting the 
scattered power-law and thermal component account for the bulk of the 
soft X-ray flux below $\sim 1$ keV, hence explaining  the lack of variability 
in the 0.1--1.0 keV band. 

It should be noted that there are many unresolved questions relating to 
both the 0.1--1 keV spectrum and the complex absorption apparent
in NGC 4151. For example, the lack of a prominent O VII 
line at 570 eV, O VIII L$_{\alpha}$ line at 650 eV and the Fe-L blend 
near 1 keV in the {\sl ASCA} spectra limits the contribution of 
thermal sources with temperature in the range $10^{6}-10^{7}$ K. 
This is already a problem for the model in Figure 8b where the observed 
equivalent width of the OVII line is $\sim 150$ eV whereas 
the {\sl ASCA \/} spectrum gives an upper limit close to 50 eV for an 
isolated narrow line feature at an energy of 570 eV. This is unfortunate
since the $10^{6}$ K thermal component can account for roughly
$\sim 25\%$ of the count rate measured by the {\sl ROSAT} HRI 
on NGC 4151, which is comparable to the fraction attributed to the extended
emission (Morse et al. 1995). Since the extended emission is very 
likely to be thermal in origin (unless the X-ray emission is strongly 
anisotropic - see Morse et al.) the lack of strong line emission below
1 keV could indicate either gas with a low metal abundance or 
gas which is over-ionized (as a result of the incident photoionizing flux) 
relative to coronal equilibrium conditions. This problem is even more 
acute if the temperature of the extended emission is  closer to $10^{7}$ K. 

\subsection{The hard X-ray/gamma-ray continuum in NGC 4151}

The earlier spectral analysis has been largely based on an assumed value
for the energy index of the hard X-ray continuum, $\alpha = 0.5$. 
Compatibility with the combined {\sl ASCA \/} SIS-0 and 
{\sl CGRO\/} OSSE spectral data is then obtained 
with the inclusion of an 
exponential cut-off term with a characteristic energy $E_{c} 
\approx 90$ keV. If the spectral index is allowed to be a free parameter 
in the original partial covering model discussed above, the resulting best 
fit gives $\alpha \approx 0.45$, in line with our assumption. 
>From the results in Table 6 we obtain an absorption-corrected flux 
in the 2--10 keV band of $3.6 \times 10^{-10} \rm~erg~cm^{-2}~s^{-1}$.
This corresponds to a particularly bright state of the source with 
reference to the compilation of results from {\sl EXOSAT} and {\sl Ginga} 
reported in Yaqoob et al. (1993). Figure 2 of Yaqoob et al. (1993) shows 
that such X-ray bright states are more typically characterised by 
$\alpha = 0.6-0.7$. However, there is no real inconsistency with the present 
measurements since the continuum slope of NGC 4151 is not particularly well 
constrained by the combination of the {\sl ASCA} and OSSE
spectra (mainly because the strong absorption significantly reduces the
effective bandwidth for the former). As a minor point we also note that
exponential form for the high energy cut-off serves to steepen the 
effective spectral slope in the medium energy band by 
$\Delta\alpha \sim 0.05$.

The thermal Comptonization model discussed in section 5 and, in the 
NGC 4151 context by Maisack et al. (1993), Titarchuk \&
Mastichiadis (1994) and Zdziarski, Johnson \& Magdziarz (1995), provides
a physical basis for the observed continuum form in this AGN. 
Specifically the model of Titarchuk \& Mastichiadis (1994) when 
fitted to the composite data (again in the context of the partial 
covering model) gives $kT = 49^{+8}_{-6}$ keV and $ \alpha = 
0.64^{+0.03}_{-0.02}$, values which are very similar to those
obtained from the OSSE only spectral fits. Here we derive
a somewhat steeper value for the asymptotic low-energy spectral index
than was obtained with the cut-off power-law model due presumably 
to differences in the exact form of the two spectral models within the 
limited spectral coverage. As emphasised by
Zdziarski, Johnson \& Magdziarz (1995), the hard spectrum of NGC 4151 
implies that the hot Comptonized plasma is starved of soft photons, 
a situation which could arise, for example, in very patchy
corona situated above the surface of an accretion disk.

The gap in spectral coverage between the {\sl  ASCA \/} 
and OSSE data prevents a very sensitive test for the presence
of a Compton reflection component (e.g. Zdziarski, Johnson \& Magdziarz 
1995). However, if we fix $\alpha = 0.9$ (a typical value for Seyfert 1 
galaxies - see Nandra \& Pounds 1994) in the cut-off power-law model 
and include a reflection component (with  $cos~i$ set to 0.45), 
then the best fit to the observed spectrum requires a relative normalisation 
for the reflection, $R = 2.2$ (see Magdziarz \& Zdziarski 1995). 
Since {\sl Ginga} data for this object strongly rule out such a strong
reflection component  (Maisack \& Yaqoob 1991; Yaqoob et al. 1993),
this in turn implies that the continuum spectrum of NGC 4151 is indeed
atypical of Seyfert 1 galaxies as a class. However, an interesting point
is that $R=2.2$ implies an equivalent width of the iron K-line
of $\sim 300$ eV (for a solar abundance reflector) which is comparable to 
the equivalent width observed when a full account is taken of both
broad and narrow line components in the 5--8 keV band  
(see Yaqoob et al. 1995 for a full discussion of the iron-line 
properties of NGC 4151).
 
\section{The short-timescale X-ray variability in NGC 4151}

The primary objective of the program discussed here (and 
in Papers I, II and IV) is the study the short time-scale 
spectral variability exhibited by NGC 4151. The derived light curves in the 
1--2 keV, 2--10 keV and 50--150 keV
bands are shown in Figures 4 and 6. Also Table 7 summarises the basic 
properties of the variability in each of these bands (we exclude the 
0.1-1.0 keV range since there is no evidence for variability in this band).
For each energy range the table lists the mean flux, the fractional 
measurement error 
(which is the average value of the flux error divided by the mean flux) 
and the fractional variability (which is the standard deviation of the 
fluxes divided by the mean flux). As noted earlier, the largest amplitude 
variations were recorded in the 1--2 keV band; however, an important
qualification is that the {\sl ASCA} 2--10 keV monitoring did not 
commence until after the onset of the significant flux increase 
observed by {\sl ROSAT} in the 1--2 keV band. Although the soft X-ray 
variability may, in principle, arise from either changes in the level of 
the underlying continuum or changes in the line-of-sight absorption, 
the {\sl ROSAT\/} data marginally favour the former possibility 
(see section 3), 
and we focus on this interpretation in this paper. The difference between
the soft X-ray and gamma-ray light curves requires a steepening of the 
continuum as the source brightens in X-rays, and suggests that the 
wide-band variability takes the form of a ``pivoting'' of the
continuum around a point near $\sim 100$ keV (see section 5; 
Yaqoob \& Warwick 1991; Zdziarski, Johnson \& Magdziarz 1995).
 
Figure 9 compares the soft X-ray 1--2 keV and the ultraviolet 
1440 \AA~ light curves for NGC 4151 (see Paper I for details of 
the {\sl IUE} measurements).
A detailed investigation of the temporal relation between the soft X-ray 
and longer wavelength variability is presented in Paper IV 
(Edelson et al. 1995) and here we restricted our attention to  
the form of the ultraviolet to X-ray correlation which is 
apparent in Figure 9. The main feature of the soft X-ray light curve, 
namely the marked flux increase which commenced roughly 3 days into 
the observation, coincides with a rather similar variation in the 
ultraviolet continuum band. However, the relative amplitude of the 
X-ray variability is significantly higher than is seen even in the 
shortest wavelength ultraviolet bands (compare the results in 
Table 7 with those given in Table 5 of Paper I).
It also apparent that the correlation between the ultraviolet 
and soft X-rays is far 
from perfect; for example, the ultraviolet flux at 1440 \AA~ shows 
a gradual decline over the second half of the 
observing period which was not mirrored in soft X-ray band. 
In order to investigate the form of the correlation between the X-ray 
and the ultraviolet continuum flux we have extracted the
1440 \AA~fluxes averaged over a period of $\pm 0.3$ d 
corresponding to each {\sl ROSAT} measurement. In addition the
observed 1--2 keV fluxes were scaled to absorption-corrected 
2--10 keV continuum fluxes, $F^{c}$(2-10 keV), based on the spectral model 
in Table 6 (thermal bremsstrahlung case). In this process the 
contribution of the (non-varying) thermal bremsstrahlung component 
is automatically removed. Figure 10 shows the resulting correlation plot.  
If we exclude the last 
{\sl ROSAT} observation (which was taken 3 days after the main group
of observations and at the time when the decline in the ultraviolet flux
had no clear correspondence in the X-ray data), then there is 
good correlation between the two wavebands (correlation coefficient = 0.86).

The X-ray/ultraviolet correlation seen in the current observations 
may be compared to that found earlier by 
Perola et al. (1986) in a comparison of {\sl EXOSAT} and {\sl IUE}
measurements (represented by the dotted line in Figure 10). 
Perola \& Piro (1994) 
have discussed the earlier data in terms of a model in which a central 
spherical source illuminates a thin accretion disk with a hard
continuum which is reprocessed to give rise to the correlated ultraviolet 
flux. In this model the heat input due to the incident radiation 
is presumed to dominate that due to viscous heating of the disk.  
This is consistent with the fact that  the X-ray/ultraviolet correlation 
observed by Perola et al. (1986) implies almost no residual 
(uncorrelated) ultraviolet flux. Perola \& Piro (1994) found that in 
order to match the observed ultraviolet flux and ultraviolet
spectral index it was necessary to assume that the gamma-ray continuum
extended out to $\sim 4$ MeV, contrary to current {\sl CGRO} OSSE
results (e.g. section 6; Zdziarski, Johnson \& Magdziarz, 1995).
However, in a recent paper Zdziarski \& Magdziarz (1995) argue
that if the source of the X-ray to gamma-ray continuum is
a patchy dissipative corona situated above the surface of the
accretion disk, then the problem of powering the observed
ultraviolet flux is largely overcome. Although the slope of the 
X-ray/ultraviolet correlation measured in the present observations
(Figure 10) is similar to that measured by Perola et al. (1986), 
a very considerable 
residual ultraviolet flux is now predicted when the 2--10 keV X-ray 
emission is extrapolated to zero.  It is possible that this
additional ultraviolet flux is evidence of an enhancement in
the viscous heating processes in the disk. However, 
in the reprocessing scenario it should be noted that although the soft X-ray 
continuum is highly variable, at gamma-ray energies, where the bulk of the 
energy resides, the continuum is much more stable. Thus an alternative
possibility is that the residual ultraviolet component arises from the 
reprocessing of the high energy end of the continuum. This would suggest
that the different correlations evident in Figure 10 mirror a significant 
change in the form of the gamma-ray spectrum of NGC 4151 between 
1986 and 1993.

\section{Conclusions}

We have presented a new analysis of the high energy properties of NGC
4151 based on {\sl ROSAT}, {\sl ASCA} and {\sl CGRO} OSSE data. 
The soft X-ray to gamma-ray
spectrum of NGC 4151 measured in simultaneous observations is dominated 
by a hard power-law continuum, which is cut-off at both high ($\sim90$ keV)
and low ($\sim4$ keV) energy. The high energy cut-off may originate
from the process of thermal Comptonization whereas that at low energy 
certainly arises from absorption in line-of-sight gas. In NGC 4151
this gas may be partially photoionized by the continuum source but still 
retains significant opacity below 1 keV. As noted in earlier studies 
there is a soft excess below 1 keV which probably arises from more 
than one scattered and/or thermal components.

Of the various X-ray and gamma-ray wavebands monitored in the
present campaign the largest amplitude of variability 
($\sim 30\%$ rms) is observed in the 1.0--2.0 keV range. 
The most obvious feature in the soft X-ray light
curve is a marked increase (factor 1.45) in the count rate over a timescale 
of $\sim 2$ days. During the same period the count rates 
recorded by OSSE 
in the 50--150 keV gamma-ray band show no more than $\pm 15\%$
changes and no clear correspondence with the soft X-ray variations.
The difference between the soft X-ray and gamma-ray light curves 
may be interpreted as a  steepening of the continuum as the source 
brightens in X-rays and suggests the wide-band variability takes the form
of a ``pivoting'' of the continuum around a point near $\sim 100$ keV.

In contrast, the 1--2 keV light curve does show a reasonable correspondence 
with that observed in the ultraviolet continuum, although the relative 
amplitude of the variations is again much higher in the X-ray data. 
A possible interpretation of the correlated ultraviolet to X-ray variability
is that the hard (X-ray to gamma-ray) continuum in NGC 4151 is produced 
in a patchy dissipative corona above the surface of an accretion disk 
and that the ultraviolet flux results from the reprocessing of part of this 
continuum by the disk. The origin of the uncorrelated (residual) 
ultraviolet flux is uncertain but reprocessing and/or the viscous 
heating of the disk are possibilities.

\vspace{1.0cm}

\section{Acknowledgements} 

This research has been supported in part by the NASA grants NAG5-2439, 
NAG5-1813, NAGW-3129 and NAS5-30960. David Smith acknowledges support from 
PPARC in the form of a research studentship.

\pagebreak

\section{References}

Crenshaw, D.M. et al., 1995, ApJ, submitted (Paper I) \\
Edelson, R.A. et al., 1995, ApJ, submitted (Paper IV) \\
Elvis, M., Briel, U.G., Henry, J.P., 1983, ApJ, 268, 105\\
Elvis, M., Fassnacht, C., Wilson, A.S., Briel, U., 1990, ApJ, 361, 459\\
Fiore, F., Perola, G.C., Romano, M., 1990, MNRAS, 243, 522\\
Holt, S.S., Mushotzky, R.F., Becker, R.H., Boldt, E.A., Serlemitsos, P.J., 
Szymkowiak, A.E., White, N.E., 1980, ApJ, 241, L13\\
Johnson, W.N., Kinzer, R.L., Kurfess, J.D., Strickman, M.S.,
Purcell, W.R., Grabelsky, D.A., Ulmer, M.P., Hillis, D.A., Jung, G.V.,
Cameron, R.A., 1993, ApJS, 86, 693 \\
Kaspi, S., et al., 1995, ApJ, submitted (Paper II) \\
Krolik, J.H., Kallman, T.R., 1987, ApJ, 320, L5\\
Lampton, M., Margon, B., Bowyer, S., 1976, ApJ, 208, 177\\
Lawrence, A., 1980, MNRAS, 192, 83\\
Magdziarz, P., Zdziarski, A. A., 1995, MNRAS, 273, 837 \\
Maisack, M., Yaqoob, T., 1991, AA, 249, 25\\
Maisack, M. et al., 1993, ApJ, 407, L61 \\
Marshall, F.E. et al., 1993, ApJ, 405, 168\\
Morse, J.A., Wilson, A.S., Elvis, M., Weaver, K.A., 1995, ApJ, 439, 121 \\
Morrison, R. McCammon, D., 1983, ApJ, 270, 119\\
Nandra, K., Pounds, K., 1994, MNRAS, 268, 405 \\
Netzer, H., 1993, ApJ, 411, 594\\
Perola, G.C. et al., 1986, ApJ, 306, 508\\
Perola, G.C, Piro, L., 1994, AA, 281, 7 \\
Smith, D.A., Done, C., 1995, MNRAS, in press \\
Snowden, S.L., Plucinsky, P.P., Briel, U., Hasinger, G.,
Pfeffermann, E., 1992, ApJ, 393, 819 \\
Tanaka, Y., Inoue, H., Holt, S.S., 1994, PASJ, 46, L37 \\
Titarchuk, L., Mastichiadis, A., 1994, ApJ, 433, L33 \\
Warwick, R.S., Done, C., Smith, D.A., 1995, MNRAS, 275, 1003 \\
Weaver, K.A., Mushotzky, R.F., Arnaud, K.A, Serlemitsos, P.J, Marshall, 
F.E., Petre, R., Jahoda, K.M., Smale, A.P., 1994a, ApJ, 423, 621 \\
Weaver, K.A., Yaqoob, T., Holt, S.S., Mushotzky, R.F., Matsuoka, M.,
Yamuchi, M., 1994b, ApJ, 436, L27  \\
Yaqoob, T., Warwick, R.S., 1991, MNRAS, 248, 773\\
Yaqoob, T., Warwick, R.S., Pounds, K.A., 1989, MNRAS, 236, 153\\
Yaqoob, T., Warwick, R.S., Makino, F., Otani, C., Sokoloski, J.L., 
Bond, I.A., Yamauchi, M., 1993, MNRAS, 262, 435 \\
Yaqoob, Y., Edelson, R., Weaver, K., Warwick, R.S., Mushotzky, R.F.,
Serlemitsos, P.J., Holt, S.S., 1995, ApJ, 453, L81 \\
Zdziarski A. A., Johnson N. W., Done C., Smith D., McNaron-Brown K., 
1995, ApJ, 438, L63 \\
Zdziarski A. A., Johnson N. W.,  Magdziarz, P. 1995, MNRAS, in press \\
Zdziarski, A.A., Magdziarz, P., 1995, MNRAS, in press \\

\pagebreak 

{\bf Figure captions}

{\bf Figure 1.} The {\sl ROSAT} PSPC light curves of NGC 4151 (filled circles)
and the BL Lac object 1E1207.9+3945 (stars) in the low (0.1 -- 0.4
keV), medium (0.5 -- 1.0 keV) and high-energy (1.0 -- 2.0 keV) {\sl ROSAT}
bands. The origin of the time axis is UT 00hr on 1993 December 1 ($=$
Julian Date 2449322.5).

{\bf Figure 2.} The PSPC pulse height spectrum for NGC 4151 from the
first six {\sl ROSAT} observations (group A dataset) ratioed to that from the 
last five observations (group B dataset).

{\bf Figure 3.} Minimum $\chi^{2}$ contours for the simultaneous
spectral fitting of the Group A and B spectra. The two axes correspond to
fixed ratios for both the normalisation $A$ of the hard power-law component
and the column density $N_{H}$ across the group A and 
B datasets. The minimum $\chi^{2}$ occurs at a point consistent with 
a normalisation change of $\sim 2.4$ but with little corresponding 
variation in $N_{H}$. The contours correspond to 68, 90 and 99 percent 
confidence ranges based on a $\Delta\chi^{2}$ criterion of 2.3, 4.6 and 
9.2, respectively.

{\bf Figure 4.} The X-ray light curve of NGC 4151. (a) The measured fluxes
in the 1--2 keV band. (b) The measured fluxes in the 2--10 keV band.
The filled circles are {\sl ROSAT} PSPC measurements and the diamonds are the
{\sl ASCA} measurements. The flux units are $10^{-12}\rm~erg~s^{-1}~cm^{2}$.
The horizontal dashed line corresponds to the contribution of the 
(non-varying) thermal bremsstrahlung component to the 1--2 keV flux.

{\bf Figure 5.} The ratio of the four {\sl ASCA} SIS-0 spectra to the best-fitting
model (the normalisation of the hard power-law is the only variable
parameter across the four data sets - see text). The top to bottom panels
correspond to {\sl ASCA\/} observations 1 to 4 respectively.

{\bf Figure 6.} The 50-150 keV light curve from {\sl CGRO\/} OSSE observations.

{\bf Figure 7.} (a) The best-fitting partial covering model (including
a thermal bremsstrahlung representation of the flux below 1 keV) 
compared to the composite {\sl ROSAT\/}--{\sl ASCA\/}--OSSE count rate 
spectra. (b) 
The spectral fitting residuals below $\sim2$ keV for the {\sl ROSAT\/} PSPC
(circled crosses) and {\sl ASCA\/} SIS 0 data (plain crosses). (c) The 
deconvolved 0.1 to $\sim 300$ keV spectrum of NGC 4151.

{\bf Figure 8.} (a) A spectral model for NGC 4151 in which a  warm absorber
replaces the partial covering. The ionization parameter has a value 
$\xi = 28$. In this case the soft flux below 1 keV is due to a 
combination of scattered and leakage flux. (b) A warm absorber model with a 
reduced value of the ionization parameter ($\xi = 14$) in which the bulk 
of the soft flux is due to a scattered power-law component and to
a soft thermal component.

{\bf Figure 9.} A comparison of the 1440 \AA~ light curve
obtained from {\sl IUE} with the 1--2 keV light curve from {\sl ROSAT} and 
{\sl ASCA}.
The dashed line in the X-ray light curve corresponds to the estimated 
contribution of the (non-varying) soft X-ray emission (see section 3).
The flux units are  $\rm10^{-14}~erg~cm^{-2}~s^{-1}~\AA^{-1}$ for
the ultraviolet bands and $10^{-12}\rm~erg~cm^{-2}~s^{-1}$ 
for the soft X-ray measurements.

{\bf Figure 10.} A comparison of the measured absorption-corrected 
flux in the 2--10 keV band with that observed at 1440 \AA~. The flux
units are $10^{-11}\rm~erg~cm^{-2}~s^{-1}$ and $\rm10^{-14}~erg~cm^{-2}~
s^{-1}~\AA^{-1}$ in the two bands respectively. 
No correction for reddening has been applied to the ultraviolet data.
The dashed line represents the best-fitting linear correlation (excluding 
one extreme point).
The dotted line is the  linear correlation derived by Perola et al. (1986)
from {\sl EXOSAT} and {\sl IUE} measurements (after scaling to the units
of the plot).

\pagebreak

\begin{table}[h]

\caption{The {\sl ROSAT\/} PSPC Observations}
\begin{center}
\begin{tabular}{ccccc}
& & & \\
UT Start & UT End & JD-2440000$^{a}$ & Exposure$^{b}$ & $F_{1-2}$~$^{c}$ \\
         &        &               &                &                  \\
         &        &               &                &                  \\
Nov-30 23:50:30  & Dec-01 00:22:45 & 9322.50 & 1317 &  $2.3^{+0.6}_{-0.6}$ \\
Dec-01 12:34:55  & Dec-01 14:33:03 & 9323.07 & 1427 &  $3.3^{+0.6}_{-0.6}$ \\
Dec-01 23:44:14  & Dec-02 00:16:01 & 9323.50 & 1217 &  $3.3^{+0.6}_{-0.6}$ \\
Dec-02 12:28:33  & Dec-02 14:26:46 & 9324.06 & 1412 &  $3.1^{+0.6}_{-0.6}$ \\
Dec-02 23:37:41  & Dec-03 00:09:32 & 9324.50 & 1145 &  $3.5^{+0.7}_{-0.7}$ \\
Dec-03 12:22:01  & Dec-03 14:20:33 & 9325.06 & 1302 &  $2.3^{+0.6}_{-0.6}$ \\
Dec-04 01:06:45  & Dec-04 01:40:32 & 9325.56 & 1192 &  $4.0^{+0.6}_{-0.6}$ \\
Dec-04 12:16:47  & Dec-04 14:13:32 & 9326.05 & 1387 &  $4.3^{+0.6}_{-0.6}$ \\
Dec-04 18:43:18  & Dec-04 21:03:49 & 9326.33 & 3084 &  $5.1^{+0.4}_{-0.4}$ \\
Dec-05 06:02:04  & Dec-05 06:37:18 & 9326.76 & 1298 &  $6.4^{+0.7}_{-0.7}$ \\
Dec-05 16:07:25  & Dec-05 23:48:46 & 9327.33 & 6374 &  $4.8^{+0.3}_{-0.3}$ \\
Dec-06 10:51:50  & Dec-06 12:21:18 & 9327.98 &  845 &  $5.3^{+0.8}_{-0.8}$ \\
Dec-10 13:59:20  & Dec-10 20:26:29 & 9332.22 & 2214 &  $5.1^{+0.5}_{-0.5}$ \\
                 &                 &        &      &                      \\

\end{tabular}
\end{center}
\raggedright

$^{a}$ Julian Date at mid-observation. \\
$^{b}$ Net exposure time in seconds. \\
$^{c}$ Observed flux in the 
1--2 keV band in units of $10^{-12}$ erg cm$^{-2}$ s$^{-1}$. \\

\end{table}

\begin{table}[h]

\caption{The {\sl ASCA\/} Observations}
\begin{center}
\begin{tabular}{cccccc}
         &        &              &                &
&                  \\
UT Start & UT End & JD-2440000$^{a}$ & Exposure$^{b}$ & $F_{1-2}$~$^{c}$ 
& $F_{2-10}$~$^{d}$ \\ 
         &        &              & & &            \\
         &        &              & & &             \\
Dec-04 08:33:30  & Dec-04 15:26:34 & 9326.00 & 10912 & $5.3\pm0.2$ & 
$23.8\pm0.7$ \\
Dec-05 21:03:46  & Dec-06 02:36:21 & 9327.49 &  8392 & $4.9\pm0.2$ & 
$21.1\pm0.7$ \\
Dec-07 11:14:09  & Dec-07 18:11:51 & 9329.11 &  9849 & $5.5\pm0.2$ & 
$25.1\pm0.7$ \\
Dec-09 20:52:25  & Dec-10 04:07:37 & 9331.52 & 12049 & $4.9\pm0.2$ & 
$21.3\pm0.7$ \\
& & & & & \\

\end{tabular}
\end{center}
\raggedright

$^{a}$ Julian Date at mid-observation. \\
$^{b}$ Net exposure time in seconds.\\
$^{c}$ Observed flux in the 
1--2 keV band in units of $10^{-12}$ erg cm$^{-2}$ s$^{-1}$. \\ 
$^{d}$ Observed flux in the 
2--10 keV band in units of $10^{-11}$ erg cm$^{-2}$ s$^{-1}$.\\

\end{table}
\begin{table}[h]

\caption{The {\sl CGRO\/} OSSE Observations}
\begin{center}
\begin{tabular}{ccccc}
& & & & \\ 
& & & & \\
UT Start & UT End & JD-2440000$^{a}$ & Exposure$^{b}$ &   
$F_{50-150}$~$^{c}$ \\
& & & & \\
& & & & \\
Dec-01 17:36:57 &  Dec-02 23:36:57  & 9323.86 & 53490 &  $3.41 \pm 0.23$ \\
Dec-03 00:37:26 &  Dec-03 22:53:45  & 9325.00 & 40810 &  $3.98 \pm 0.27$ \\
Dec-04 00:01:26 &  Dec-04 23:38:24  & 9326.00 & 44325 &  $4.10 \pm 0.26$ \\
Dec-05 00:43:12 &  Dec-05 22:56:38  & 9327.00 & 41140 &  $4.53 \pm 0.27$ \\
Dec-06 00:00:00 &  Dec-06 23:44:09  & 9328.00 & 44700 &  $3.56 \pm 0.26$ \\
Dec-07 02:25:26 &  Dec-07 22:59:31  & 9329.03 & 37960 &  $4.17 \pm 0.28$\\
Dec-08 00:00:00 &  Dec-08 23:44:09  & 9330.00 & 42140 &  $4.21 \pm 0.26$ \\
Dec-09 00:47:31 &  Dec-09 23:00:57  & 9331.00 & 41800 &  $4.56 \pm 0.26$\\
Dec-10 00:02:52 &  Dec-10 23:48:28  & 9332.00 & 40440 &  $4.07 \pm 0.27$ \\
Dec-11 00:51:50 &  Dec-11 23:03:50  & 9333.00 & 37690 &  $4.83 \pm 0.28$ \\
Dec-12 00:04:19 &  Dec-12 23:49:55  & 9334.00 & 40910 &  $4.06 \pm 0.27$ \\
Dec-13 00:53:16 &  Dec-13 13:49:26  & 9334.81 & 23240 &  $4.70 \pm 0.36$ \\
& & & & \\

\end{tabular}
\end{center}
\raggedright

$^{a}$ Julian Date at mid-observation. \\
$^{b}$ Net exposure time in seconds. \\
$^{c}$ Observed flux in the 50-150 keV band in units of
$10^{-11}$ erg cm$^{-2}$ s$^{-1}$.\\

\end{table}

\begin{table}[h]
\caption{Spectral fitting of the {\sl ROSAT\/} PSPC data}

\begin{center}
\begin{tabular}{cccc}
& & & \\
Parameter & Group A/B &  Group A & Group B \\
          & Datasets     &  Dataset & Dataset \\      
& & & \\
$A^{a}$           &   -     & $1.0^{+0.6}_{-0.3}$  & $2.4^{+0.5}_{-0.4}$ \\
$\alpha$          & $0.5$  &       -               &       -               \\
$N_{H}$$^{b}$     &   -     & $225^{+70}_{-50}$    & $240^{+30}_{-30}$   \\
& & & \\
$A_{brem}$$^{a}$    & $1.4^{+0.3}_{-0.3}$           & - & - \\
$kT_{brem}$$^{c}$   & $0.44^{+0.05}_{-0.04}$        & - & - \\
$N_{Hgal}$$^{b}$    & $2.6^{+0.3}_{-0.2}$           & - & - \\
& & & \\
$\chi^{2}$ & 59.4 & & \\
d-o-f      & 47   & & \\
& & & \\
\end{tabular}
\end{center}
\raggedright

$^{a}$ $10^{-2}$ photon cm$^{-2}$ s$^{-1}$ keV$^{-1}$. \\
$^{b}$ $10^{20}$ cm$^{-2}$. \\
$^{c}$ keV. \\

\end{table}

\begin{table}[h]
\caption{The partial covering model applied to the {\sl ASCA\/} data}

\begin{tabular}{cccccc}
& & & & & \\
Parameter & \multicolumn{5}{c} {Observation} \\                     
&  All & 1 & 2 & 3 & 4  \\
& & & & & \\
$A^{a}$              & - & $6.5^{+0.2}_{-0.2}$ & $5.8^{+0.2}_{-0.2}$ &
$6.9^{+0.2}_{-0.2}$ & $5.8^{+0.2}_{-0.2}$ \\
$\alpha$             & $0.5$ & - & - & - & - \\
& & & & & \\
$N_{H1}$$^{d}$       & $1120^{+370}_{-270}$ & - & - & - & - \\
$C_{F}$              & $0.42^{+0.06}_{-0.05}$ & - & - & - & -  \\
$N_{H2}$$^{d}$       & $350^{+30}_{-30}$ & - & - & - & - \\ 
& & & & & \\
$E_{K\alpha}$$^{c}$  & $6.34^{+0.03}_{-0.04}$ & - & - & - & - \\
$I_{K\alpha}$$^{b}$  & $2.9^{+0.6}_{-0.6}$ & - & - & - & - \\
& & & & & \\
$A_{brem}$$^{a}$     & $0.30^{+0.06}_{-0.05}$ & - & - & - & - \\
$kT_{brem}$$^{c}$    & $1.3^{+0.3}_{-0.2}$  & - & - & - & - \\
$N_{Hgal}$$^{d} $    & $2.1$   & - & - & - & - \\
& & & & &\\
$\chi^{2}$ & 1371 & & & & \\
d-o-f      & 1052 & & & & \\
& & \\
& & & & & \\

\end{tabular}
\raggedright

$^{a}$ $10^{-2}$ photon cm$^{-2}$ s$^{-1}$ keV$^{-1}$. \\
$^{b}$ $10^{-4}$ photon cm$^{-2}$ s$^{-1}$.\\
$^{c}$ keV.\\
$^{d}$ $10^{20}$ cm$^{-2}$.\\
\end{table}

\begin{table}[h]
\caption{The partial covering model applied to the {\sl ROSAT\/}-{\sl
ASCA}-{\sl CGRO} data}

\begin{tabular}{ccc}
& & \\
Parameter   & Thermal Brems.   & Power-law     \\
            & Soft Excess      & Soft Excess \\
& & \\
$A^{a}$                   & $6.5^{+0.2}_{-0.2}$ & $6.6^{+0.2}_{-0.2}$  \\
$\alpha$                  & $0.5$   & $0.5$    \\
$E_{c}$$^{c}$             & $94^{+4}_{-4}$ & $93^{+4}_{-4}$  \\
  			  &         &         \\
$N_{H1}$$^{d}$            & $540^{+80}_{-50}$ & $630^{+310}_{-140}$ \\
$C_{F}$                   & $0.75^{+0.07}_{-0.08}$ & $0.57^{+0.14}_{-0.15}$ \\
$N_{H2}$$^{d}$            & $210^{+40}_{-40}$ & $320^{+50}_{-60}$ \\
                          &         &         \\ 
$A_{brem}$/$A_{soft}$$^{b}$ & $9.7^{+0.9}_{-0.8}$ & $1.86^{+0.10}_{-0.11}$  \\
$\alpha_{soft}$           & - & $1.3^{+0.2}_{-0.2}$  \\
$kT_{brem}$$^{c}$         & $0.51^{+0.03}_{-0.04}$  & -       \\
$N_{Hgal}$$^{d}$          & $2.1 < 2.2$   & $2.2^{+0.2}_{-0.1}$   \\
& &\\
$\chi^{2}$ & 388 & 378 \\
d-o-f      & 300 & 300 \\
& & \\
\end{tabular}
\raggedright

$^{a}$ $10^{-2}$ photon cm$^{-2}$ s$^{-1}$ keV$^{-1}$. \\
$^{b}$ $10^{-3}$ photon cm$^{-2}$ s$^{-1}$ keV$^{-1}$. \\
$^{c}$ keV. \\
$^{d}$ $10^{20}$ cm$^{-2}$. \\

\end{table}









\begin{table}[h]

\caption{The Variability Parameters}
\begin{center}
\begin{tabular}{lccc}
         &    &   &  \\
Band     & Mean         & Fractional  & Fractional        \\
         & Flux~$^{a}$  & Error$^{a}$ & Variability$^{c}$ \\  
         &              &             &                   \\
1--2 keV~$^{b}$    &  0.41   & 0.06    &  0.31  \\
2--10 keV          &  22.8   & 0.03    &  0.09  \\
50-150 keV         &  4.18   & 0.06    &  0.10  \\
& & & \\

\end{tabular}
\end{center}
\raggedright

$^{a}$ Observed flux in the band in units of 
$10^{-11}$ erg cm$^{-2}$ s$^{-1}$. \\ 
$^{b}$ Based only on the {\sl ROSAT} observations. \\
$^{c}$ {\it Not} corrected for the statistical errors.

\end{table}

\pagestyle{empty}

\newpage

\begin{figure}
\vspace{17cm}
\includegraphics{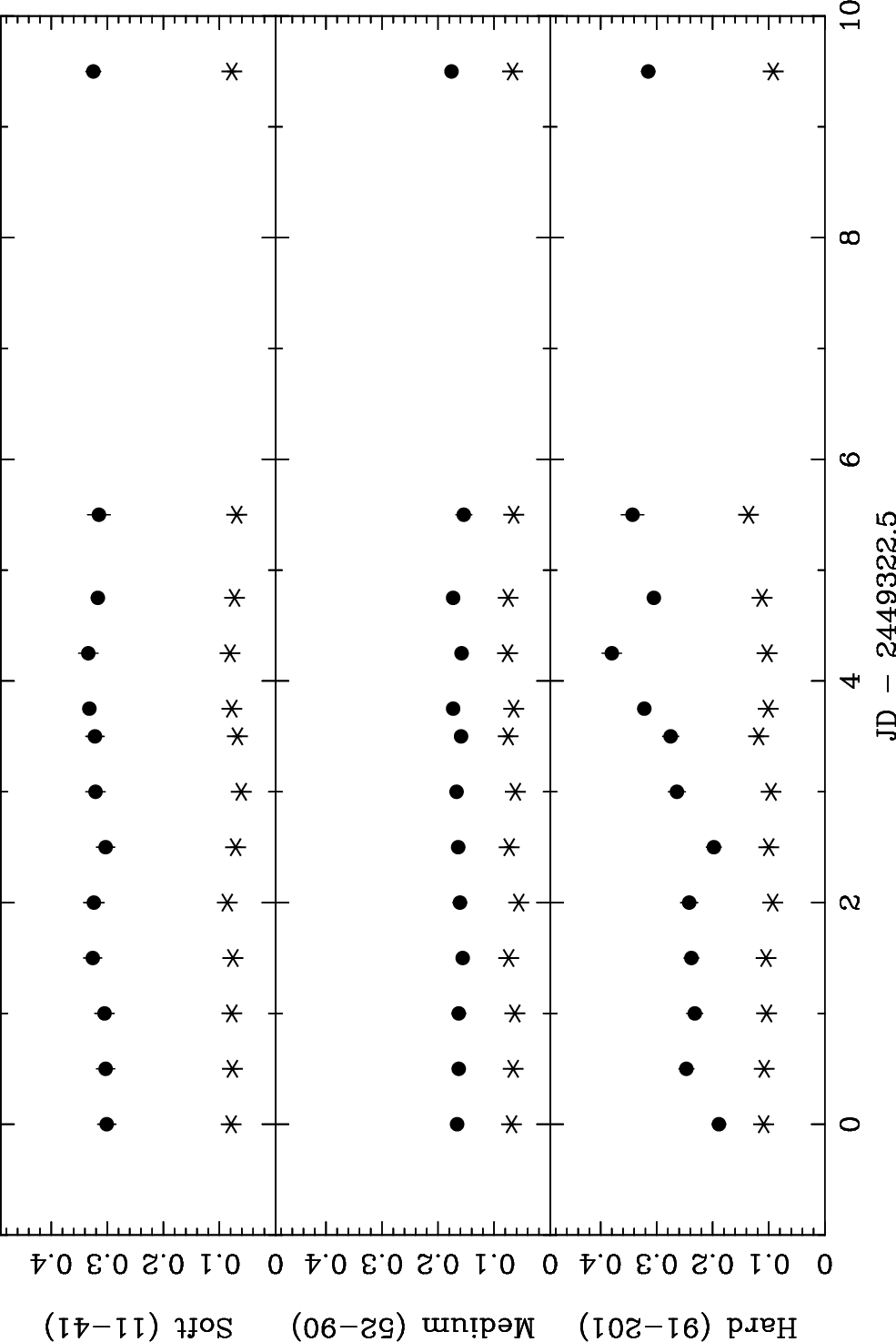}
\caption{ }
\end{figure}

\begin{figure}
\vspace{10cm}
\includegraphics{p3fig2.ps}
\caption{ }
\end{figure}

\begin{figure}
\vspace{10cm}
\includegraphics{p3fig3.ps}
\caption{ }
\end{figure}

\begin{figure}
\vspace{13cm}
\includegraphics{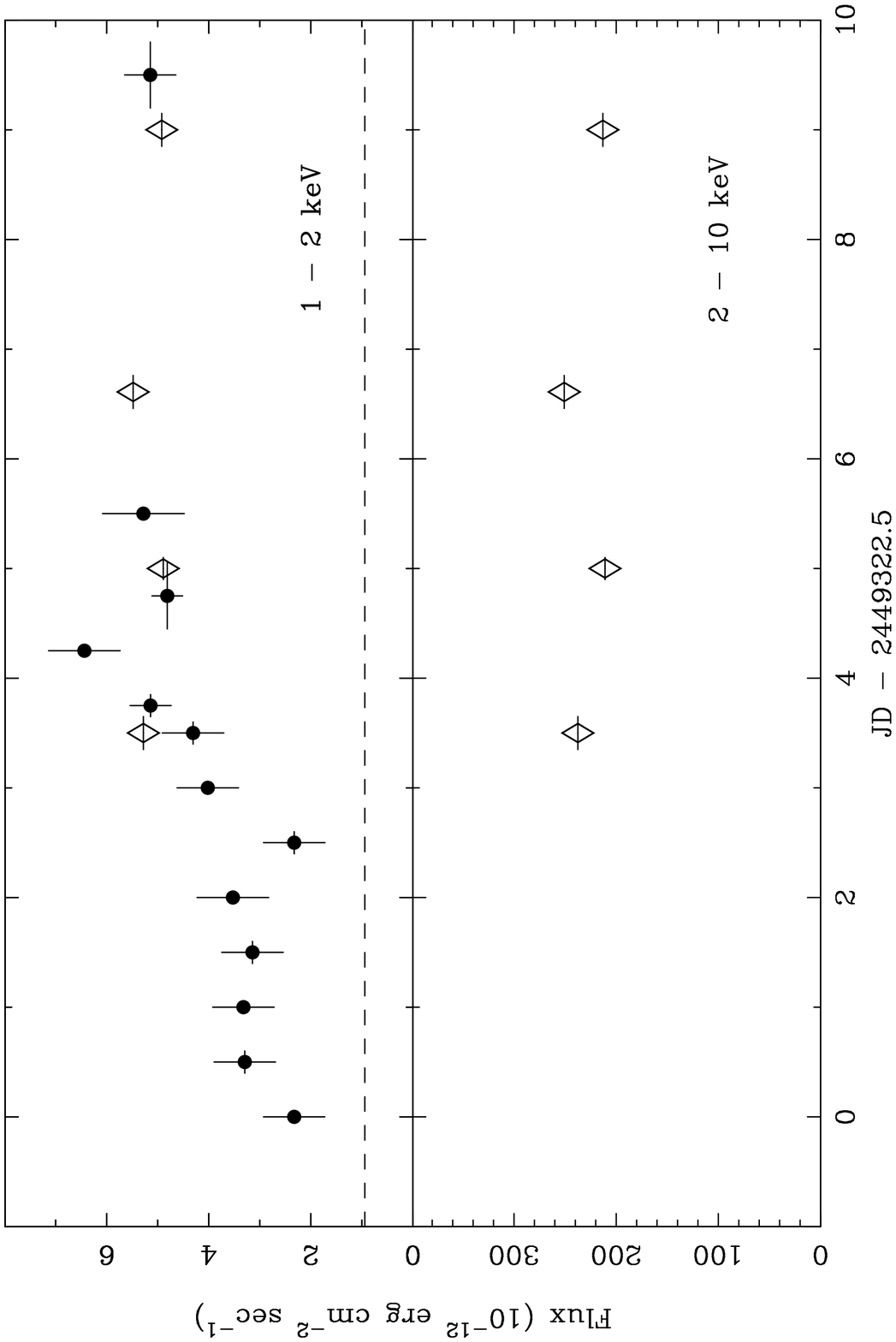}
\caption{ }
\end{figure}

\begin{figure}
\vspace{11cm}
\includegraphics{p3fig5.ps}
\caption{ }
\end{figure}

\begin{figure}
\vspace{10cm}
\includegraphics{p3fig6.ps}
\caption{ }
\end{figure}

\begin{figure}
\vspace{24cm}
\includegraphics{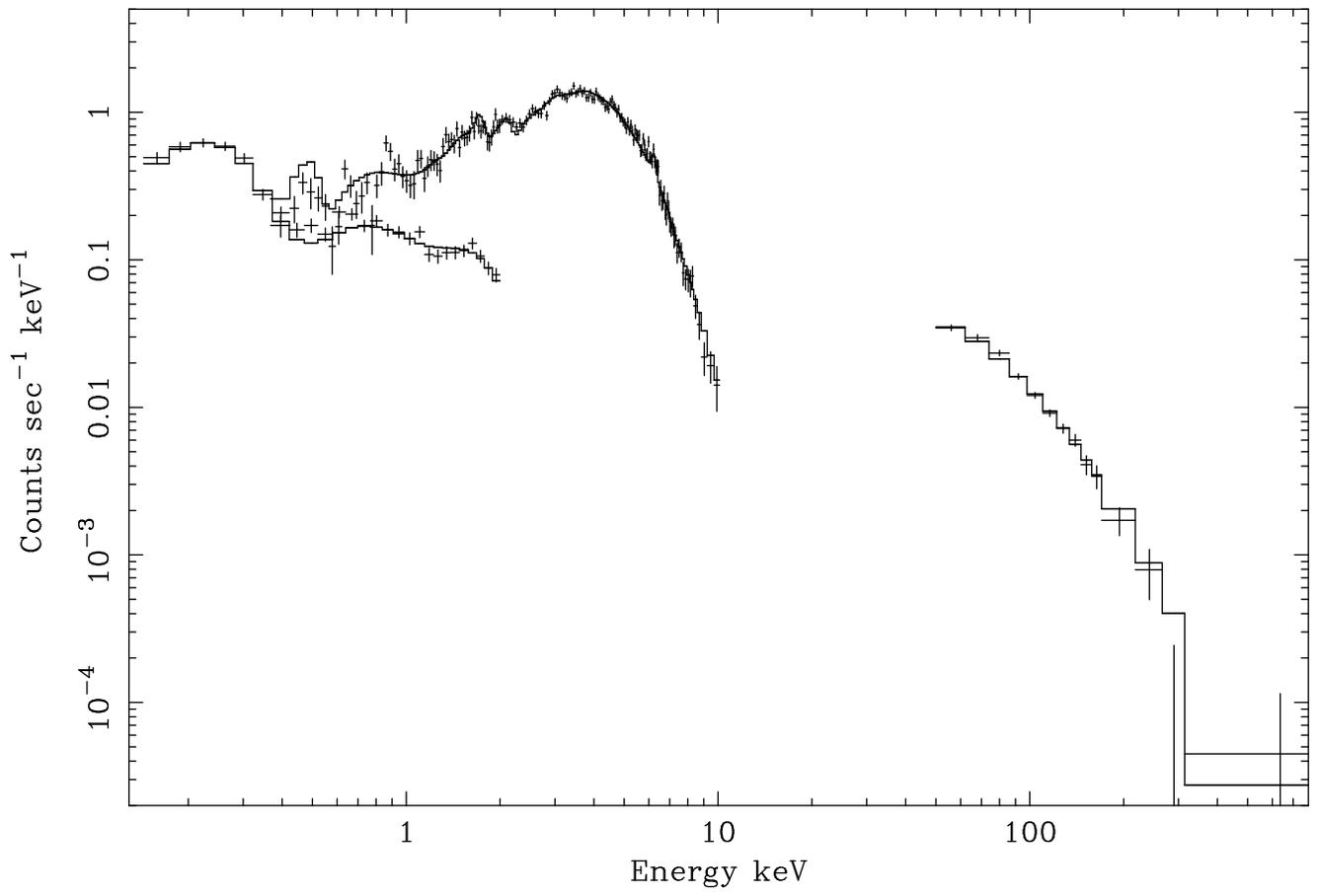}
\includegraphics{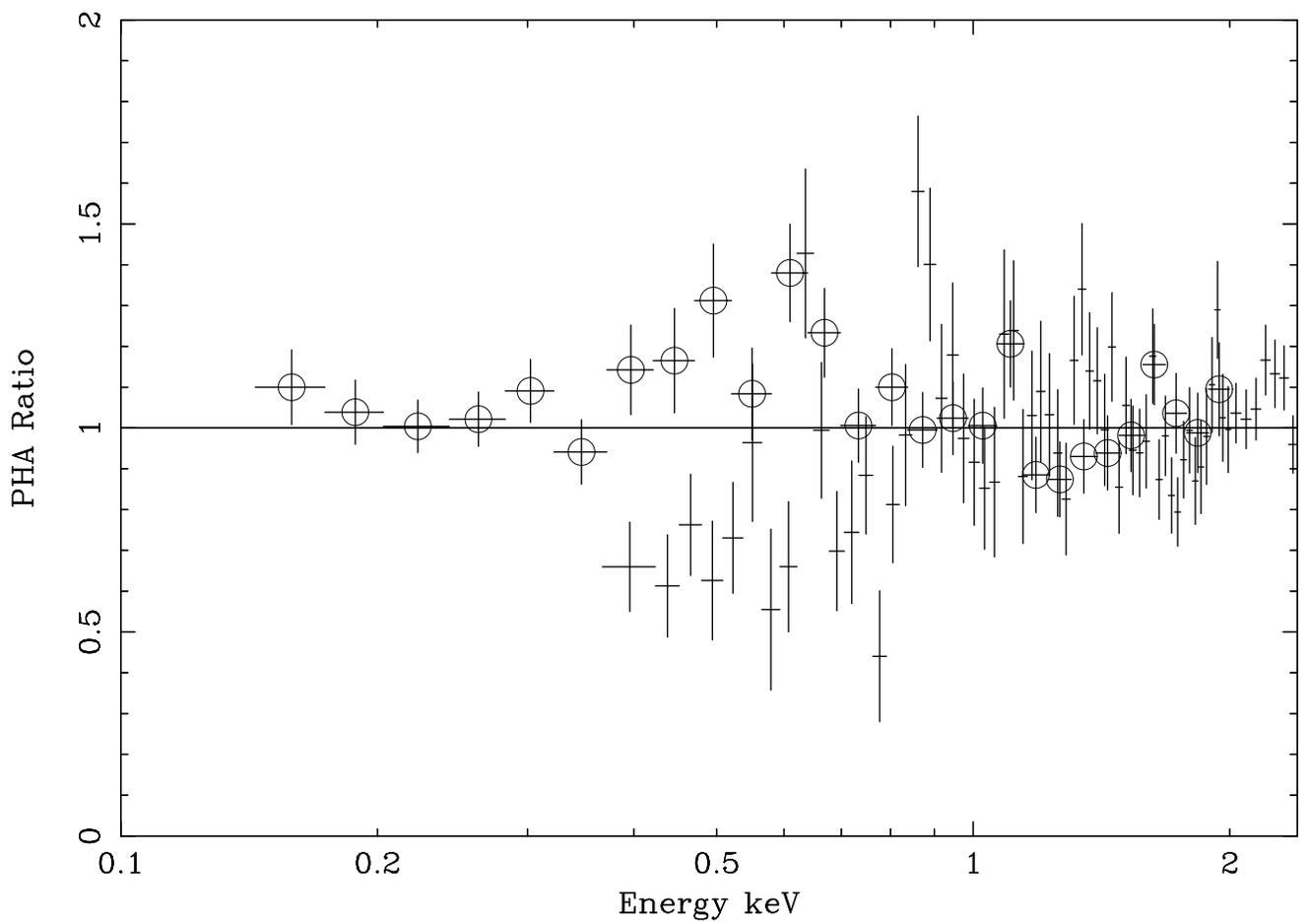}
\caption{a - top, b - bottom, c - next page}
\end{figure}

\begin{figure}
\vspace{17cm}
\includegraphics{p3fig7c.ps}
\end{figure}

\begin{figure}
\vspace{24cm}
\includegraphics{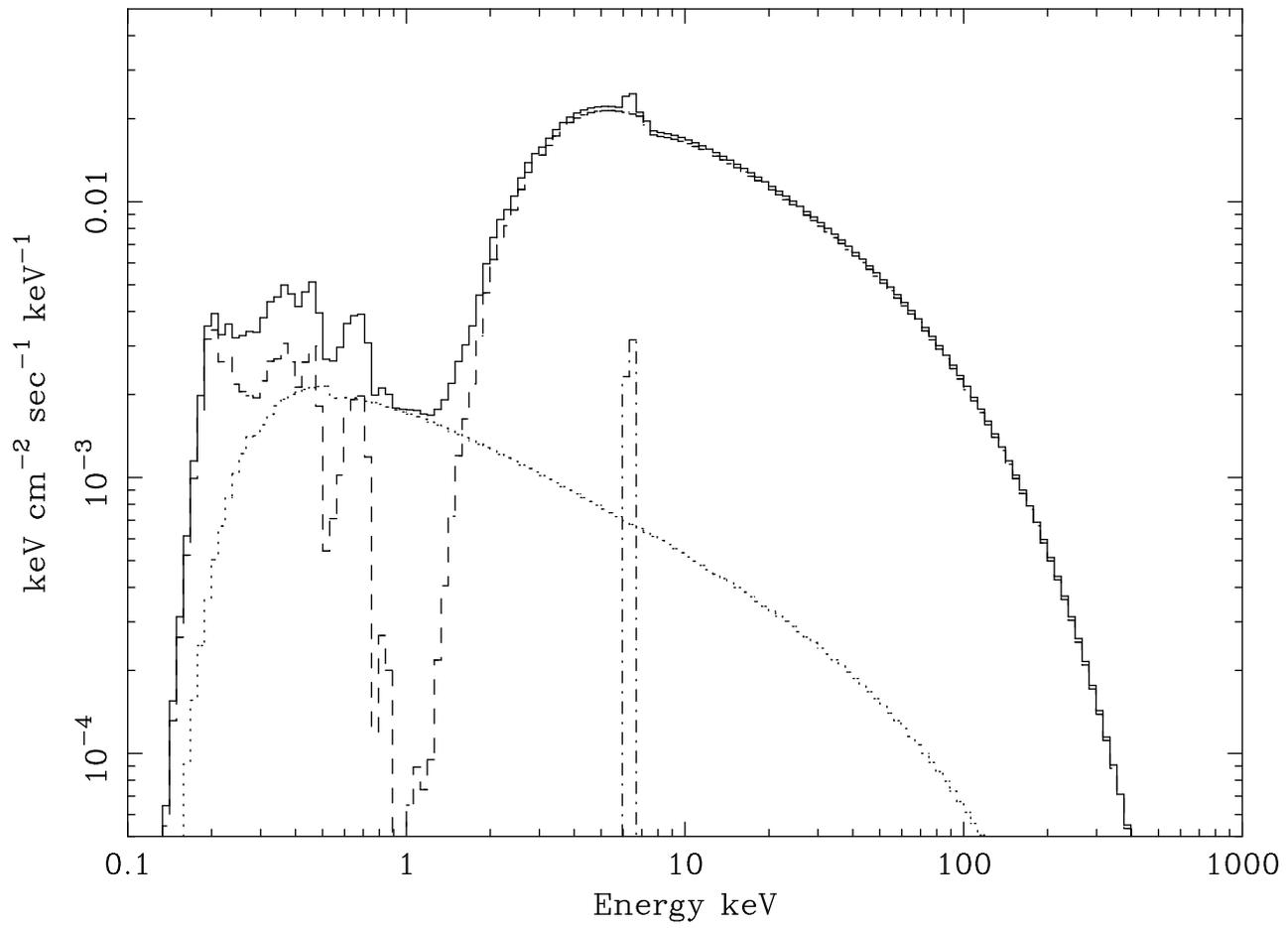}
\includegraphics{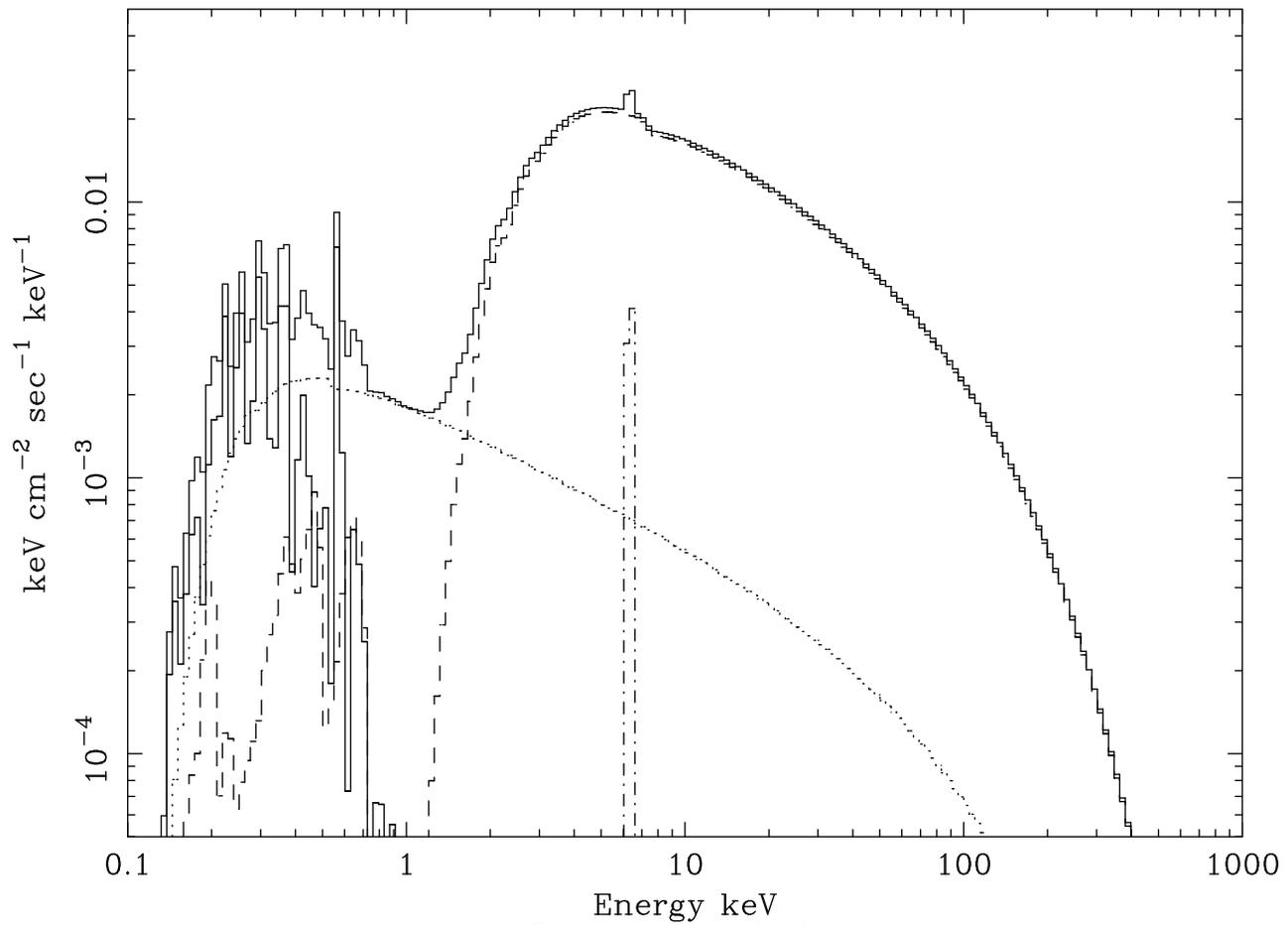}
\caption{a - top, b - bottom}
\end{figure}

\begin{figure}
\vspace{13cm}
\includegraphics{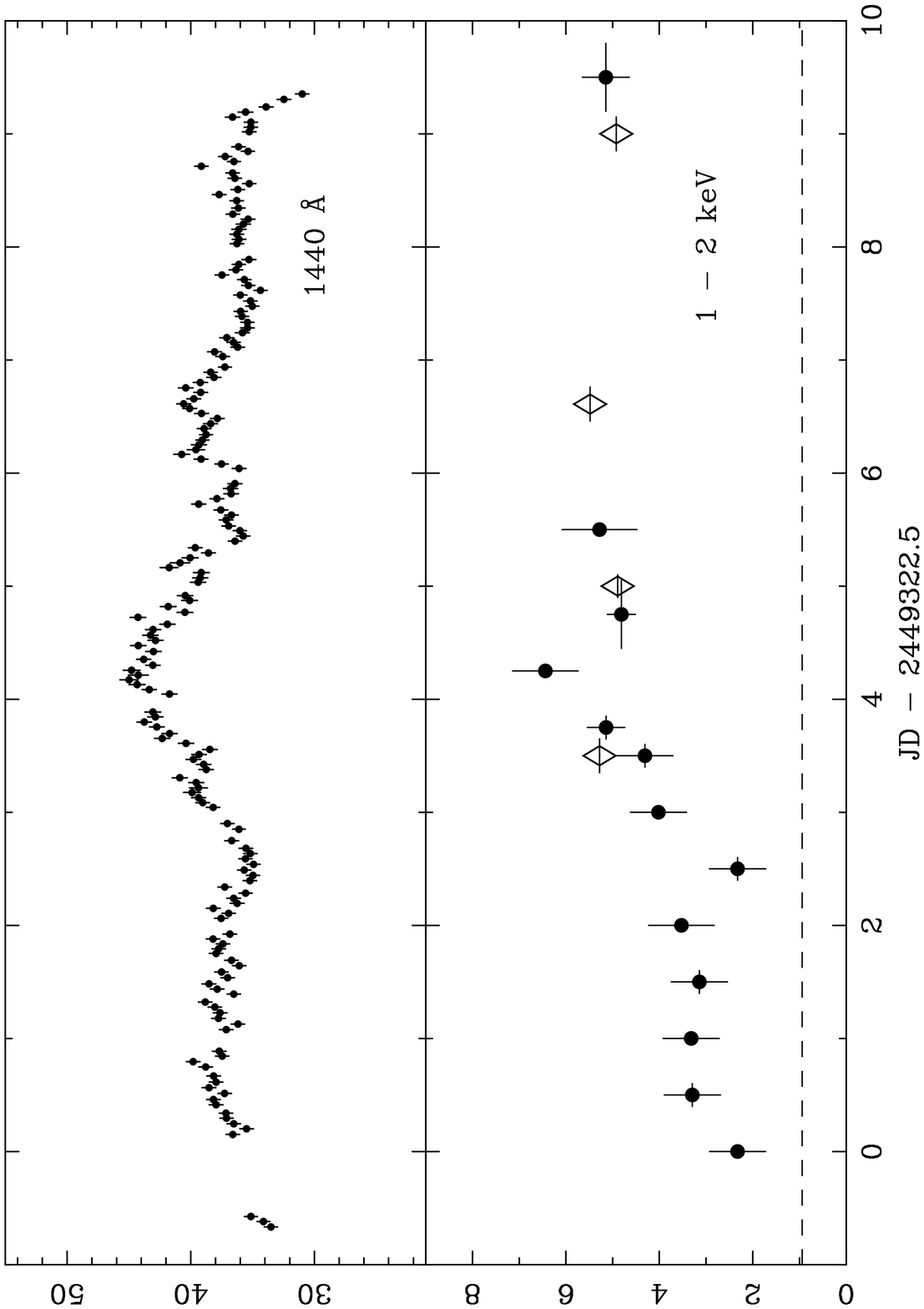}
\caption{ }
\end{figure}

\begin{figure}
\vspace{15cm}
\includegraphics{p3fig10.ps}
\caption{ }
\end{figure}

\end{document}